\documentclass[prd]{revtex4}
\usepackage{subfig}
\usepackage{bm} 
\usepackage{graphicx,epsfig}
\usepackage{latexsym,amssymb,amsmath,float,url}
\usepackage{latexsym}
\usepackage{cancel}
\usepackage{color}

\begin{document}

\title{Damping of Primordial Gravitational Waves from Generalized Sources}

\author{James B. Dent}
\affiliation{Physics Department, University of Louisiana-Lafayette, Lafayette, LA, 70503}

\author{Lawrence M. Krauss}
\affiliation{School of Earth and Space Exploration and Department of Physics, Arizona State University, Tempe, AZ 85287 and \\Research School of Astronomy and Astrophysics, Australian National University, Canberra, Australia 2614}

\author{Subir Sabharwal, and Tanmay Vachaspati}
\affiliation{Department of Physics, Arizona State University, Tempe, AZ 85287}

\begin{abstract}
It has been shown that a cosmological background with an anisotropic stress tensor, appropriate for a free streaming  thermal neutrino background, can damp primordial gravitational waves after they enter the horizon, and can thus affect the CMB B-mode polarization signature due to such tensor modes.  Here we generalize this result, and examine the sensitivity of this effect to non-zero neutrino masses, extra neutrino species, and also a possible relativistic background of axions from axion strings.  In particular, additional neutrinos with cosmologically interesting neutrino masses at the $O(1)$ eV level will noticeably reduce damping compared to massless neutrinos for gravitational wave modes with $k\tau_0 \approx 100-200$, where $\tau_0 \approx 2/H_0$ and $H_0$ is the present Hubble parameter, while an axion background would produce a phase-dependent damping distinct from that produced by neutrinos.
\end{abstract}

\maketitle

\section{Introduction}

A generic prediction of inflation in the early universe \cite{Guth:1930zm,Linde:1981mu,Albrecht:1982wi} is the production of 
gravitational waves (GW) with a nearly flat spectrum \cite{Grishchuk:1974ny,Starobinsky:1979ty}.  
There are ongoing observational efforts to detect such a spectrum and recent, comprehensive 
reviews of past, present, and future experimental efforts can be found in 
Refs.~\cite{Pryke:2012mi, Krauss:2010ty}.  
The dominant signature in the near term involves the effect that long-wavelength gravitational waves 
can have on the cosmic microwave background (CMB) through the generation of B-mode 
polarization ({\it e.g.}, 
\cite{Kosowsky:1994cy,Kamionkowski:1996ks,Zaldarriaga:1996xe,Hu:1997hp,Hu:1997hv,
Cabella:2004mk,Baumann:2008aq}).  The amplitude of the gravitational waves can be related to the energy 
scale at which inflation occurred, and the ratio of the power spectrum of gravitational waves to that of 
the scalar power spectrum, also known as the ``tensor-to-scalar ratio'', $r$, can give vital information 
into the nature of the inflaton -- the field which drives inflation -- via the Lyth bound 
\cite{Lyth:1996im}.  Additionally, primordial gravitational wave spectra produced by various non-inflationary 
mechanisms have been suggested 
\cite{Krauss:1991qu,JonesSmith:2007ne,Fenu:2009qf,Krauss:2010df}.  Therefore any 
observation of a primordial gravitational wave spectrum 
would be an immensely powerful tool in the study of the very early universe.

A question that naturally arises is - are there any effects that can intervene and alter the nature of the GW spectrum from the time of its production until the time of observation?  If the answer is yes, then one must account for such effects in order to accurately describe the primordial spectrum.  As is well known, just such an intervening effect does arise due to the fact that an anisotropic stress from free streaming particles can damp the amplitude of GWs from their primordial value.  Weinberg showed in \cite{Weinberg:2003ur} that the damping effect of free-streaming neutrinos on the GW spectrum can be quite significant with up to 35.6\% loss in amplitude, and following this work, the issue has been the subject of some attention \cite{Bashinsky:2005tv,Dicus:2005rh,Boyle:2005se,Watanabe:2006qe,Miao:2007cw,Mangilli:2008bw,Zhao:2009we,Benini:2010zz,Stefanek:2012hj,Shchedrin:2012sp,Jinno:2012xb}.

The original study of Weinberg, and much of the following work has been focused on the effects of three massless neutrinos.  However, recent cosmological observations have shown hints of deviations from the standard cosmological value of three effective neutrino degrees of freedom \cite{Izotov,Komatsu,Dunkley,Keisler,Archidiacono}.  Due to neutrino oscillation experiments, it is also known that neutrinos are not massless, as described in a recent global analysis of neutrino properties \cite{Fogli:2012ua}). There are also some, albeit statistically insignificant presently, hints that the addition of extra neutrino species can improve fits to short baseline neutrino oscillation data \cite{Conrad:2012qt}. Recently issues regarding light sterile neutrinos and cosmology have been addressed in \cite{Ho:2012br,Jacques:2013xr,Mirizzi:2013kva,Archidiacono:2013fha}.    

Prompted by these results, we broaden the scope by calculating GW damping in more general scenarios, including the effect of neutrino masses, additional (beyond three) massive and massless neutrino species, and extra \emph{bosonic} degrees of freedom.  (Note that other recent work \cite{Jinno:2012xb} also touches on some of these issues related to extra degrees of freedom and gravitational waves.)  

This paper is organized as follows.  In Sec.~\ref{calculation} we derive in some generality the formula for anisotropic stress.  The results for particular cases are presented in Sec.~\ref{results}, and we then present our conclusions in Sec.~\ref{conclusions}.

\section{Calculation}\label{calculation}

To first order, a perturbed FRW metric with scale factor $a(t)$ can be written as 
\begin{eqnarray}
ds^2 = -(1+2\psi)dt^2 + a(\partial F_i + G_i)dx^idt + a^2(t)[\delta_{ij}(1 - 2\phi) + h_{ij} + \partial_i\partial_jB + \partial_jC_i + \partial_iC_j]dx^idx^j \label{1}
\end{eqnarray}
Gravitational waves arise as the transverse, traceless components of the metric fluctuations, which are characterized by $h_{ij}$.  These modes satisfy the transverse and traceless conditions
\begin{eqnarray}
\partial^ih_{ij} = 0 \;\;\;\;; \;\;\;\; h_{ii} = 0 \label{2}
\end{eqnarray}
The Fourier transformed $k$-space modes, $h_k$ satisfy the Einstein equation of the form
\begin{eqnarray}
h_k'' + 2Hh_k' + k^2h_k = 16\pi G_N a^2(\tau) \Pi_k \label{3}
\end{eqnarray}
where the prime denotes differentiation with respect to conformal time $d\tau = dt/a(t)$, and $\Pi_k$ is the anisotropic stress.  

In order to solve this equation, we first turn to the Boltzmann equation, which determines the evolution of the phase space density of the particles, $F(x,P)$, given as a function of the four-momentum $P$ which has components $P^{\mu} = dx^{\mu}/d\lambda$.   One can then determine the anisotropic stress by perturbing the distribution function about the background as $F(x,P) = \bar{F}(P^0) + \delta F(x,P)$, and employing the Boltzmann equation $dF(x,P)/dt = 0$.  For the scenario of three massless neutrinos, the details of this calculation have been nicely presented in Appendix D of \cite{Watanabe:2006qe},  which explores the impact of collisionless damping.  We will follow this treatment, but we will generalize to the case of massive particles with the number of degrees of freedom left as an input parameter.  We will also examine the situation where a bosonic degree of freedom is incorporated, as well as the case of relativistic axions produced by axionic cosmic strings. These relativistic axions have a non-thermal spectrum. 

For particles with a thermal distribution, the background phase space density is given by
\begin{eqnarray}
\bar{F}(P^0) = \frac{g}{e^{P^0/T_{\nu}}\pm1}\label{4}
\end{eqnarray}
where the plus sign is for fermions while the minus sign is for bosons, and
$g$ gives the number of degrees of freedom. 
For the two cases of interest in this work: $g = g_{\nu} = 2$ for a single neutrino, and $g = g_B =1$ for a real scalar.
$T_{\nu}$ is the temperature of neutrinos which is related to the photon temperature at times after neutrinos decoupling as $T_{\nu}=(4/11)^{1/3}T_{\gamma}$. One begins with the relation
\begin{eqnarray}
g_{\mu\nu}P^{\mu}P^{\nu} = -(P^0)^2 +g_{ij}P^iP^j = -m^2
\end{eqnarray}
We  write this as
\begin{eqnarray}\label{6}
\tilde{p}_0^2 = g_{ij}P^iP^j
\end{eqnarray}
where we have defined a new variable through a shift
\begin{eqnarray}
P^0 \equiv \sqrt{m^2 + \tilde{p}_0^2}.
\end{eqnarray}

\subsection{Damping from neutrinos}
\label{subsec:neutrinos}

We follow the derivation in \cite{Watanabe:2006qe} with this change ($\mbox{their}\ P_{0}=\mbox{our}\ \tilde{p}_{0}$). The full neutrino distribution function satisfies the relativistic collisionless Boltzmann equation,
\begin{eqnarray}
\frac{dF(t,x^{i},\gamma^{i},\tilde{p}_0)}{dt}=\frac{\partial F}{\partial t}+\frac{dx^{i}}{dt}\frac{\partial F}{\partial x^{i}}+\frac{d \tilde{p}_0}{dt}\frac{\partial F}{\partial \tilde{p}_0}+\frac{d\gamma^{i}}{dt}\frac{\partial F}{\partial\gamma^{i}}=0\label{9}
\end{eqnarray}

Using the variable $\tilde{p}$, one finds that to first order that Eq.~(\ref{9}) becomes the Einstein-Vlasov equation
\begin{eqnarray}
\left(\frac{\partial F}{\partial t}\right)_{{first\ order}}=\frac{\partial\delta F}{\partial t}+\frac{\gamma_{i}\tilde{p}_{0}}{a\sqrt{m^2+\tilde{p}_{0}^2}}\frac{\partial\delta F}{\partial x^{i}}-\frac{\dot{a}}{a}\frac{(m^2+\tilde{p}_{0}^2)}{\tilde{p}_{0}}\frac{\partial\delta F}{\partial\tilde{p}_{0}}-\frac{1}{2}\frac{\partial\bar{F}}{\partial\tilde{p}_{0}}\tilde{p}_{0}\frac{\partial h_{ij}}{\partial t}\gamma^{i}\gamma^{j}=0
\end{eqnarray}
and $\gamma_i=\gamma^{i}$ are directional cosines. 

Defining $\mu\equiv\gamma^{i}k_{i}/k$ and using the spherical mode decomposition of $h_{ij}$ and $\delta F$
\begin{eqnarray}
h_{ij}=\sum_{\lambda=+,\times}\int\frac{d^3 k}{(2\pi)^3}h_{\lambda,k}(t)Q^{\lambda}_{ij}(\vec{x})
\end{eqnarray}
\begin{eqnarray}
\delta F=\sum_{\lambda=+,\times}\int\frac{d^3 k}{(2\pi)^3}f_{\lambda,k}(t,\tilde{p}_{0},\mu)\gamma^{i}\gamma^{j}Q^{\lambda}_{ij}(\vec{x})\label{16}
\end{eqnarray}
where  $Q^{\lambda}_{ij}$ are symmetric, traceless and divergenceless tensors that satisfy: $Q^{\lambda}_{ij}=Q^{\lambda}_{ji}$, ${Q^{\lambda}_{ij;a}}^{;a}(\vec{x})+k^2Q^{\lambda}_{ij}(\vec{x})=0$ and ${Q_{ij}^{\lambda}}^{;j}=0$. The covariant derivative is with respect to 
the unperturbed spatial FRW metric. Thus, the first order Einstein-Vlasov equation in terms of the 
decomposed (spherical) mode becomes
\begin{eqnarray}
\frac{\partial f_{k}}{\partial t}+\frac{ik\mu}{a}\left(\frac{\tilde{p}_{0}}{\sqrt{m^2+\tilde{p}_{0}^2}}\right)f_{k}-\frac{\dot{a}}{a}\left(\frac{m^2+\tilde{p}_{0}^2}{\tilde{p}_{0}}\right)\frac{\partial f_{k}}{\partial\tilde{p}_{0}}=\frac{1}{2}\tilde{p}_{0}\frac{\partial\bar{F}}{\partial\tilde{p}_{0}}\frac{\partial h_{k}}{\partial t}\label{17}
\end{eqnarray}

Once again, following \cite{Watanabe:2006qe} we define new variables $q^{\mu}=aP^{\mu}$ and $q^{0}=aP^{0}\equiv q$ and conformal time, $d\tau=dt/a(t)$. Then Eq.~(\ref{17}) can be written as
\begin{eqnarray}
\frac{\partial f_{k}}{\partial\tau}+\frac{ik\mu\tilde{p}_{0}}{\sqrt{m^2+\tilde{p}_{0}^2}}f_{k}=\left(\frac{q^2-a^2m^2}{q}\right)\frac{\partial\bar{F}}{\partial q}\frac{1}{2}\frac{\partial h_{k}}{\partial\tau}\label{18}
\end{eqnarray}
This equation determines the time evolution of the perturbation of distribution function $\delta F$ which in turn determines the anisotropic stress part of the perturbed energy-momentum tensor that goes into the RHS of Eq.~(\ref{3}). 
\begin{eqnarray}
\delta\mathcal{T}_{ij}=a^2\sum_{\lambda=+,\times}\int\frac{d^3 k}{(2\pi)^3}\Pi_{\lambda,k}Q^{\lambda}_{ij}(\vec{x})\label{19}
\end{eqnarray}
and
\begin{eqnarray}
\mathcal{T}_{ij}=\frac{1}{\sqrt{-g}}\int\frac{d^3 q}{(2\pi)^3q}q_{i}q_{j}F(q)\implies \delta\mathcal{T}_{ij}=\frac{1}{\sqrt{-g}}\int\frac{d^3 q}{(2\pi)^3q}{q}_{i}{q}_{j}\delta F(q)\label{20}
\end{eqnarray}
Using Eqs.(\ref{16}), (\ref{19}) and (\ref{20}), one finds that the anisotropic stress is
\begin{eqnarray}
\Pi_{\lambda,k}Q_{ij}^{\lambda}(\vec{x})=a^{-4}\int\frac{d^3q}{(2\pi)^3q}q^2\gamma^{i}\gamma^{j}
\gamma^{l}\gamma^{m}f_{\lambda,k}Q_{lm}^{\lambda}(\vec{x})\label{21}
\end{eqnarray}
where $f_{\lambda,k}\equiv f_{\lambda,k}(\tau,q,\mu)$. On the other hand, Eq.~(\ref{18}), which is a first order differential equation, has the following solution
\begin{eqnarray}
f_{k}(\tau,q,m,\mu)=\frac{q^2}{2}\frac{\partial\bar{F}}{\partial q}\int_{\tau_{dec}}^{\tau}d\tau'h'_{k}(\tau')\alpha(m,\tau',q)^2e^{-i\mu\alpha^2k(\tau-\tau')}\label{22}
\end{eqnarray}
where we have defined
\begin{eqnarray}
\alpha(\tau,m,q)^2=1-\frac{m^2a(\tau)^2}{q^2}\label{alpha}
\label{alphadefn}
\end{eqnarray}
and
\begin{eqnarray}
\mu=\frac{\gamma^{i}k_{i}}{k}\implies \mu=\hat{\gamma}\cdot\hat{k}
\end{eqnarray}
and used the fact that $f_{k}(\tau_{dec},q,m,\mu)=0$ because there is no anisotropic stress at neutrino decoupling since the neutrinos just start to free-stream at decoupling. The polarization index $\lambda$ is suppressed on both sides of Eqs.(\ref{22}). Finally using the identity
 \begin{eqnarray}
 \int d\Omega_{q}\gamma^{i}\gamma^{j}\gamma^{l}\gamma^{m}e^{-i{\gamma_{i}}{k^{i}} u}Q_{lm}^{\lambda}=\frac{1}{8}(\delta^{il}\delta^{jm}+\delta^{im}\delta^{jl})\int d\Omega_{q} e^{-i\mu u};\ \ \ d^{3}q=q^2 dq\ d\Omega_{q}
 \end{eqnarray}
one can write the anisotropic stress in momentum space (again the polarization index $\lambda$ is suppressed) 
\begin{eqnarray}
\Pi_k = \frac{1}{8a(\tau)^4}\int d\tau'\frac{d^3q}{(2\pi)^3}(1-\mu^2)^2e^{-i\mu b}h_k'(\tau')\frac{\partial \bar{F}(q)}{\partial q}q^2\alpha^2
\end{eqnarray}
where
\begin{eqnarray}
b &\equiv& k(\tau-\tau')\alpha^2\\
\alpha^2 &\equiv& 1-\frac{m^2a(\tau')^2}{q^2}\\
\mu &=& \hat{\gamma}\cdot \hat {k} = \rm{cos}\,\theta_q
\end{eqnarray}
where we have taken $\hat{k}^i$ to be in the $z$-direction in $q$-space.  We also define
\begin{eqnarray}
u &\equiv& k\tau\\
s &\equiv& k\tau'
\end{eqnarray}

We can perform the integrations over $d\Omega_q = d\phi_q d(\rm{cos}\theta_q)$ and find that anisotropic stress is then given by
\begin{eqnarray}\nonumber
\Pi_k &=& \frac{\pi}{2a(u)^4}\int_{u_{dec}}^{u} dsdq\frac{dh_k(s)}{ds}\frac{\partial \bar{F}(q)}{\partial q}q^4\alpha^2\bigg[\frac{{\rm{sin}}b}{b} +2\alpha^2\left(-\frac{{\rm{sin}}b}{b}-2\frac{{\rm{cos}}b}{b^2}+2\frac{{\rm{sin}}b}{b^3}\right)\\ &+&\alpha^4\left(\frac{{\rm{sin}}b}{b}+4\frac{{\rm{cos}}b}{b^2}-12\frac{{\rm{sin}}b}{b^3}-24\frac{{\rm{cos}}b}{b^4}+24\frac{{\rm{sin}}b}{b^5}\right)\bigg]
\end{eqnarray}

Furhermore we define
\begin{eqnarray}
x \equiv \frac{q}{aT} = \frac{q}{a_0T_0} = \frac{q}{T_0}\label{x}
\end{eqnarray}
where the second equality holds for our normalization that the present day scale factor is $a_0 = 1$.  This allows us to write the distribution $\bar{F}(q)$ and the function $\alpha$ as
\begin{eqnarray}
\bar{F}(x) = \frac{g_{\nu}}{e^{x}+1}\\
\alpha^2(m,x) = 1 - \frac{m^2a^2}{T_0^2x^2}
\end{eqnarray}

With this we find
\begin{eqnarray}\label{stress}
\Pi_k &=& \frac{1}{16\pi^2a(u)^4}\int_{u_{dec}}^{u} dsdx\frac{dh_k(s)}{ds}\frac{\partial \bar{F}(x)}{\partial x}x^4T_0^4\alpha^2(m,x)\bigg[\frac{{\rm{sin}}((u-s)\alpha^2(m,x))}{(u-s)\alpha^2(m,x)} \\\nonumber &+&2\alpha^2(m,x)\left(-\frac{{\rm{sin}}((u-s)\alpha^2(m,x))}{(u-s)\alpha^2(m,x)}-2\frac{{\rm{cos}}((u-s)\alpha^2(m,x))}{((u-s)\alpha^2(m,x))^2}+2\frac{{\rm{sin}}((u-s)\alpha^2(m,x))}{((u-s)\alpha^2(m,x))^3}\right)\\\nonumber &+&\alpha^4(m,x)\bigg(\frac{{\rm{sin}}((u-s)\alpha^2(m,x))}{(u-s)\alpha^2(m,x)}+4\frac{{\rm{cos}}((u-s)\alpha^2(m,x))}{((u-s)\alpha^2(m,x))^2}-12\frac{{\rm{sin}}((u-s)\alpha^2(m,x))}{((u-s)\alpha^2(m,x))^3}\\\nonumber &-&24\frac{{\rm{cos}}((u-s)\alpha^2(m,x))}{((u-s)\alpha^2(m,x))^4}+24\frac{{\rm{sin}}((u-s)\alpha^2(m,x))}{((u-s)\alpha^2(m,x))^5}\bigg)\bigg]
\end{eqnarray}

The full gravitational wave equation is
\begin{eqnarray}\label{wave}
\frac{d^2h_k(u)}{du^2} + 2\left(\frac{da(u)/du}{a(u)}\right)\frac{dh_k(u)}{du} + h_k(u) = \frac{16\pi G_N a^2(u)}{k^2}\Pi_k = \frac{6}{\rho(u)}\left(\frac{da(u)/du}{a(u)}\right)^2\Pi_k
\end{eqnarray}

This gives an equation for the transverse-traceless tensor modes as a general function of the mass of the particle creating the anisotropic stress for a general phase space distribution $\bar{F}(x)$.  It can be seen to reduce to the standard form for three massless neutrinos when $g_{\nu} = 6$ and $m = 0$ \cite{Watanabe:2006qe}.\\

From the relation Eq.~(\ref{stress}), one can then include additional degrees of freedom by simply using $g_{\nu} = 8$ ($10$) for four (five) massless neutrino species. We
use the simplifying assumption that the neutrinos all have the same decoupling temperature.
(To generalize to arbitrary decoupling temperatures we would change the lower integration limit 
of the anisotropic stress for that species.)  
For a mixed scenario where particles of different masses contribute, one can use 
Eq.~(\ref{stress}) for the 
anisotropic stress, $\Pi_{k\,,i}$, generated by a single species of mass $m= m_i$ with $g_{\nu} = g_i$ 
degrees of freedom, and then add another anisotropic stress term of this form for any additional 
species of mass $m_j$ with degrees of freedom $g_j$.  In other words, the total anisotropic stress 
due to $i$ particles is given by the sum $\Pi_{k, tot} = \sum_i \Pi_{k, i} $.\\

To graphically display the effect of adding non-zero neutrino masses we adopt the simple analytic form for the scale factor $a(\tau)$ in a matter plus radiation universe given by
\begin{eqnarray}
a(\tau)=\left(\frac{\tau}{\tau_{0}}\right)^2+2\left(\frac{\tau}{\tau_{0}}\right)\sqrt{a_{\rm eq}};\ \tau_{0}=\frac{2}{\sqrt{\Omega_{M}}H_{0}}\label{scalefactor}.
\end{eqnarray}
Note that today, the relation between radiation and matter densities is given by
\begin{eqnarray}
\Omega_{r}=a_{\rm eq}\Omega_{M}
\end{eqnarray}
For the standard cosmological scenario with $N_{\rm eff}=3$, {\it i.e.}, three effective 
neutrino degrees
of freedom, we have $a_{\rm eq}=1/3600$, $\Omega_{M}=0.3$ and 
$\Omega_{r}=\Omega_{\gamma}+\Omega_{\nu}$ since the free-streaming neutrinos are relativistic. 
Further it can be shown that \cite{Watanabe:2006qe}\\
\begin{eqnarray}\label{43}
\frac{\Omega_{\nu}}{\Omega_{\nu}+\Omega_{\gamma}}=0.40523;\ \ g_{\nu}=6
\end{eqnarray}
When adding extra neutrino species, we use the above relation and keep $\Omega_{M}$ fixed but change the ratio in Eq.~(\ref{43}) accordingly to get a new redshift for matter-radiation equality \cite{5yrWMAP}. So, for $N_{\rm eff}=4,\ a_{\rm eq}=1/3172$ and for $N_{\rm eff}=5,\ a_{\rm eq}=1/2834$.\\

One sees that Eq.~(\ref{wave}) is an integro-differential equation since the source term on the right-hand-side is the integral in Eq.~(\ref{stress}).  To put the equation into a suitable form for a numerical solution, we adopt the method in Appendix A of \cite{Pritchard:2004qp}, which consists of rewriting the single, second order integro-differential equation as a system of coupled first-order Volterra type integro-differential equations.  This can then be solved by standard methods of numerical integration \cite{recipes}. There is a slight difference in our method of solution from that in \cite{Watanabe:2006qe} and \cite{Pritchard:2004qp} due to the form of the integral kernel.  Namely, we do not have the simplifying option of integrating out the distribution function (which would give the energy density in the standard case of massless neutrinos) due to the additional $x$-dependence of other factors in the integrand.  We therefore were forced to generate numerical values for the integrand at each value of $u$ and $s$, after which the procedure of \cite{Pritchard:2004qp} could be implemented.

\subsection{Damping from axions}
\label{subsec:axions}

The fine tuning of $\theta_{\rm QCD}$ can be avoided in models of particle physics that contain an extra $U(1)$ Peccei-Quinn (PQ) symmetry. A consequence of these models is a light pseudo-scalar particle, the axion. In a cosmological setting, the axion is massless above the QCD temperature but gains a small mass below this temperature. Even though the axion is very light, with a typical mass $m_a =\mathcal{O}(10^{-3}~{\rm eV})$, it can be non-relativistic because it is produced coherently throughout the cosmological horizon and has momenta given by $\sim t^{-1}$ at cosmic time $t$. This argument, however, ignores the topology that accompanies the breaking of the PQ symmetry, which is relevant if the PQ symmetry breaking scale occurs below the scale of inflation. In this case, the spontaneous breaking of the PQ symmetry leads to the production of axionic cosmic strings with energy density set by the PQ energy scale. As the strings oscillate, they radiate relativistic axions. At the QCD temperature scale, the strings get connected by axionic domain walls, and the whole network of strings and walls collapses, dissipating energy again into {\it relativistic} axions. Hence the axion density in the universe contains two separate components: the non-relativistic component due to coherent oscillations of the axion field, and the relativistic component due to the radiation from topological defects. The latter component can be significant, and may even dominate the non-relativistic component for large values of the Peccei-Quinn symmetry breaking scale. Relativistic axions can also have anisotropic stress and hence they can couple to gravitational waves just as neutrinos do.

The spectral energy density of relativistic axions has been debated and there has been some disagreement between Davis \cite{Davis:1986} and Sikivie et al. \cite{Sikivie:2001}. We will be using Davis's spectral distribution which has recently been verified by field theoretic lattice simulations in \cite{Yamaguchi:1999} and \cite{Hiramatsu:2011},
\begin{eqnarray}
\rho_{a}(t)=\frac{4\pi f_{a}^2}{t^2}\int_{\Omega^{2/3}/\tilde{t}}^{\Omega^{5/6}/\sqrt{\tilde{t}t^{*}}}\left(\frac{q^2+(t/\tilde{t})m_{a}^2}{q^2+\tilde{m}^2}\right)^{1/2}\left[\ln\left(\frac{\Omega^{5/3}}{\tilde{t}q^2\delta}\right)-\frac{1}{2}\right]\frac{dq}{q}\label{axionspectrum}
\end{eqnarray}
where, as previously defined, $q$ is the comoving momentum, $f_{a}\leq 2\times 10^{10}$ GeV is the PQ symmetry breaking scale, $\delta=1/f_{a}$, $\Omega=2\pi$, $\tilde{m}\equiv1/\tilde{t}=10^{-9}-10^{-8}$ eV and $t^{*}$ is the time at which the axions decoupled. The mass of the axion $m_{a}$ and the decoupling temperature $T_{d}^{*}$ of axions is related to the scale $f_{a}$ through \cite{Sikivie:2006}
\begin{eqnarray}
m_{a}=6 \times 10^{-6} \mbox{eV}\left(\frac{10^{12}\mbox{GeV}}{f_{a}}\right)=\frac{6\times10^{15}\mbox{eV}^2}{f_{a}}
\end{eqnarray}
\begin{eqnarray}
T_{d}^{*}=5\times 10^{11}\ \mbox{GeV}\ \left(\frac{f_{a}}{10^{12}\ \mbox{GeV}}\right)^2
\end{eqnarray}
Since $T\propto 1/a(t)$, we can find $t^{*}$ using $T_{d}^{*}$ and the scale factor $a(t)$ in (\ref{scalefactor}). We will consider damping from relativistic axions with the spectrum (\ref{axionspectrum}) for three different $f_{a}$ values $10^{8}$, $10^{9}$ and $10^{10}$ GeV. This lies within the range $10^{7}\mbox{GeV}<f_{a}<2\times10^{10}\mbox{GeV}$ where the lower bound on $f_{a}$ comes from astrophysical constraints \cite{Axionfa:1987} and the upper bound which comes from the requirement that the energy density in relativistic axions remain below critical energy density to avoid overclosing the universe. 

To calculate the anisotropic stress $\Pi_{k}$, we need the unperturbed phase space distribution function $\bar{F}(q)$ of these relativistic axions which we can read off from (\ref{axionspectrum})
\begin{eqnarray}\label{eqaxion}
\bar{F}^{\rm axion}(q)=
\frac{f_{a}^2a(t)^4}{t^2q^3({q^2+a(t)^2m_{a}^2})^{1/2}}\left(\frac{q^2+(t/\tilde{t})m_{a}^2}{q^2+\tilde{m}^2}\right)^{1/2}\left[\ln\left(\frac{\Omega^{5/3}}{\tilde{t}q^2\delta}\right)-\frac{1}{2}\right]
\end{eqnarray}
However, there are few differences from the neutrino case of (\ref{stress}). Since axions have a non-thermal spectrum, we don't do the substitution of (\ref{x}) and retain the expression for $\alpha^2$ in (\ref{alpha}). Thus, the expression for $\Pi_{k}$ for axions becomes

\begin{equation}
\Pi_k^{\rm axion} = \Pi_k({\bar F} \to {\bar F}^{\rm axion})
\end{equation}
where $\Pi_k$ is defined in Eq.~(\ref{stress}) and $\alpha(m,q,\tau)^2$ is the same as in Eq.~(\ref{alphadefn})

\section{Results and Discussion}\label{results}

For massless neutrinos, the effects of damping are determined solely by the neutrino energy density contribution, 
(which falls as $a^{-4}$) once one enters the matter-dominated era, 
and will thus be most significant for $k \gtrsim k_{\rm eq} = a_{\rm eq}H_{\rm eq} \approx 170/\tau_0$.

The effect of non-zero neutrino masses will be to add an extra $k$-dependence to damping, as free streaming, and hence damping, will be reduced when the temperature is of order of the mass.  $k$-modes that come inside the horizon while neutrinos are relativistic, and contribute significantly to the overall energy density, will be damped more. On the other hand, those modes that come inside the horizon at later times, either when neutrino masses become significant, or during matter domination, when the neutrino energy density fraction may have been reduced considerably due to redshifting will be damped less.  This heuristic behavior is validated by our detailed calculations, which quantitatively explore this effect.  For demonstrations purposes here, we display the damping as a function of neutrino mass for three different values of $k\tau_0 = 100, 200,1000$.  


\begin{figure} 
\includegraphics[width=3.5in]{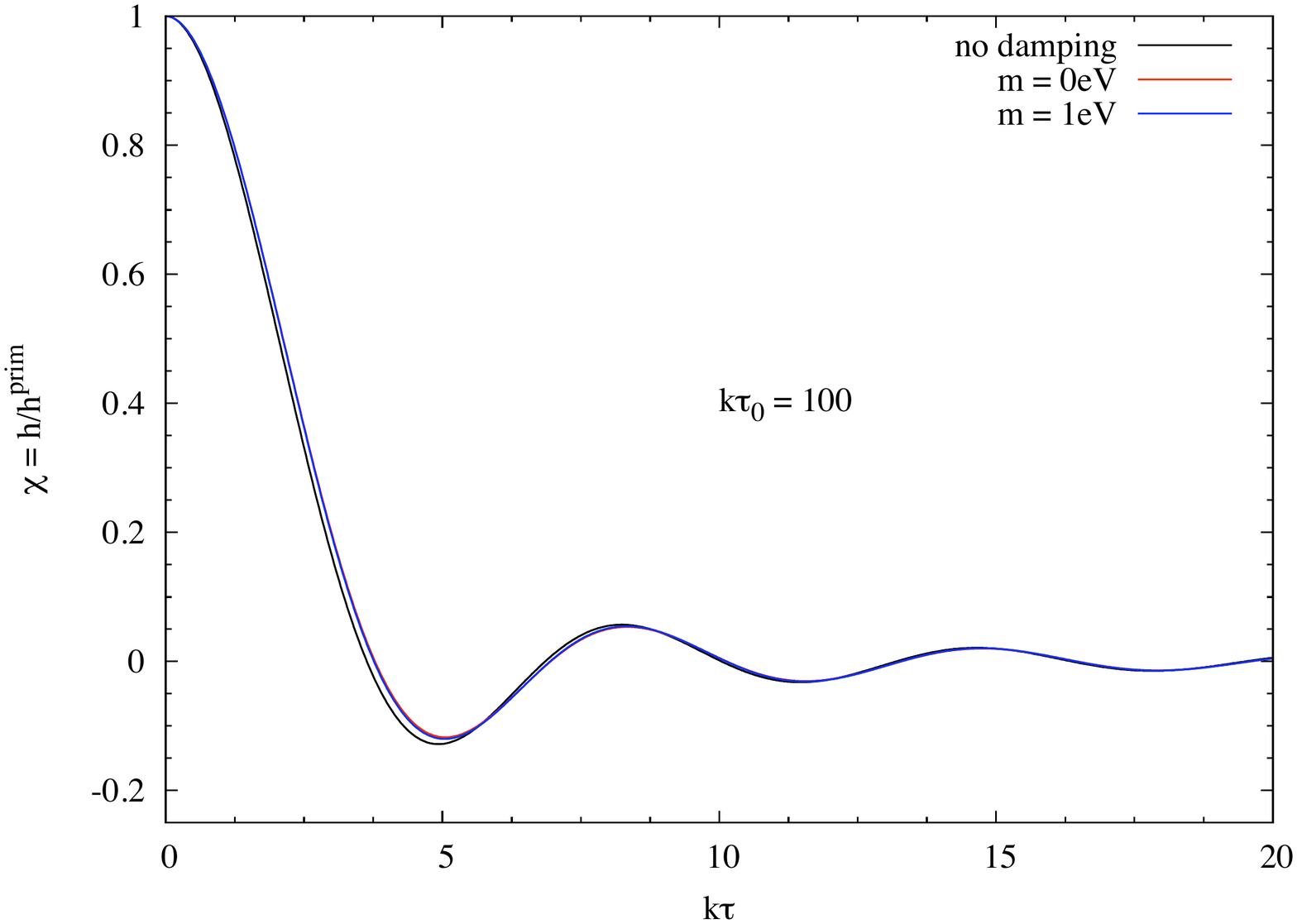} \includegraphics[width=3.5in]{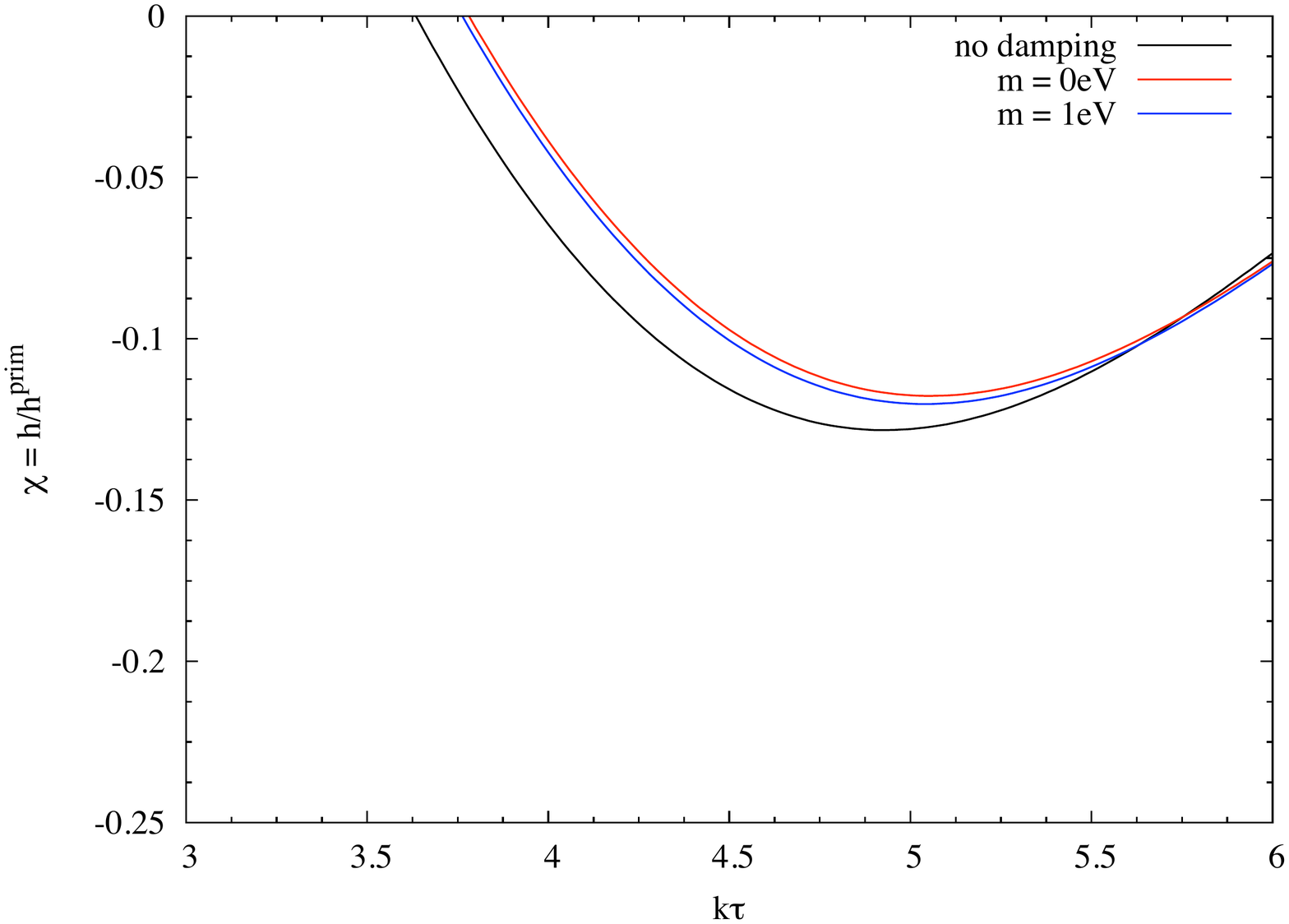}   
\includegraphics[width=3.5in]{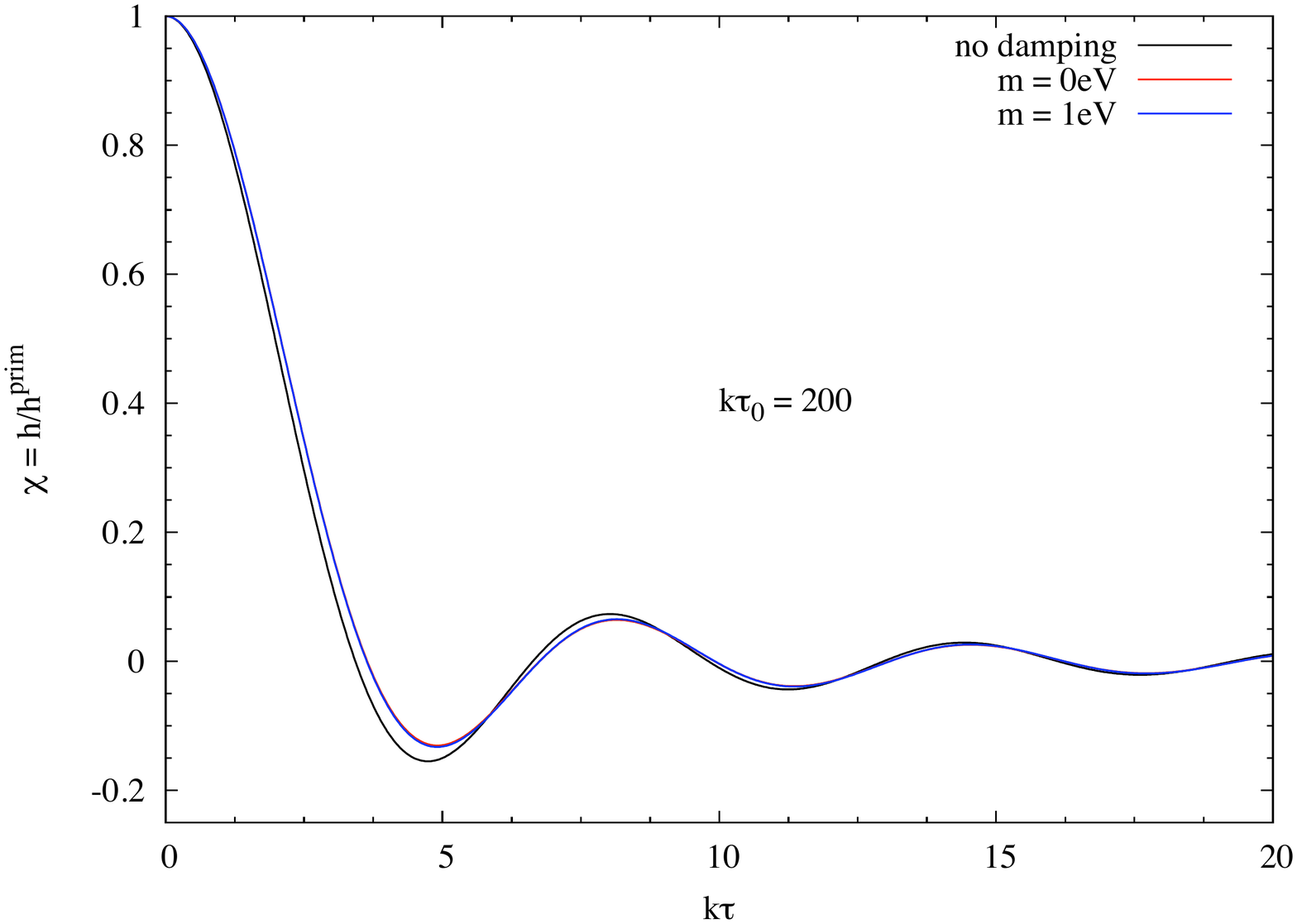} \includegraphics[width=3.5in]{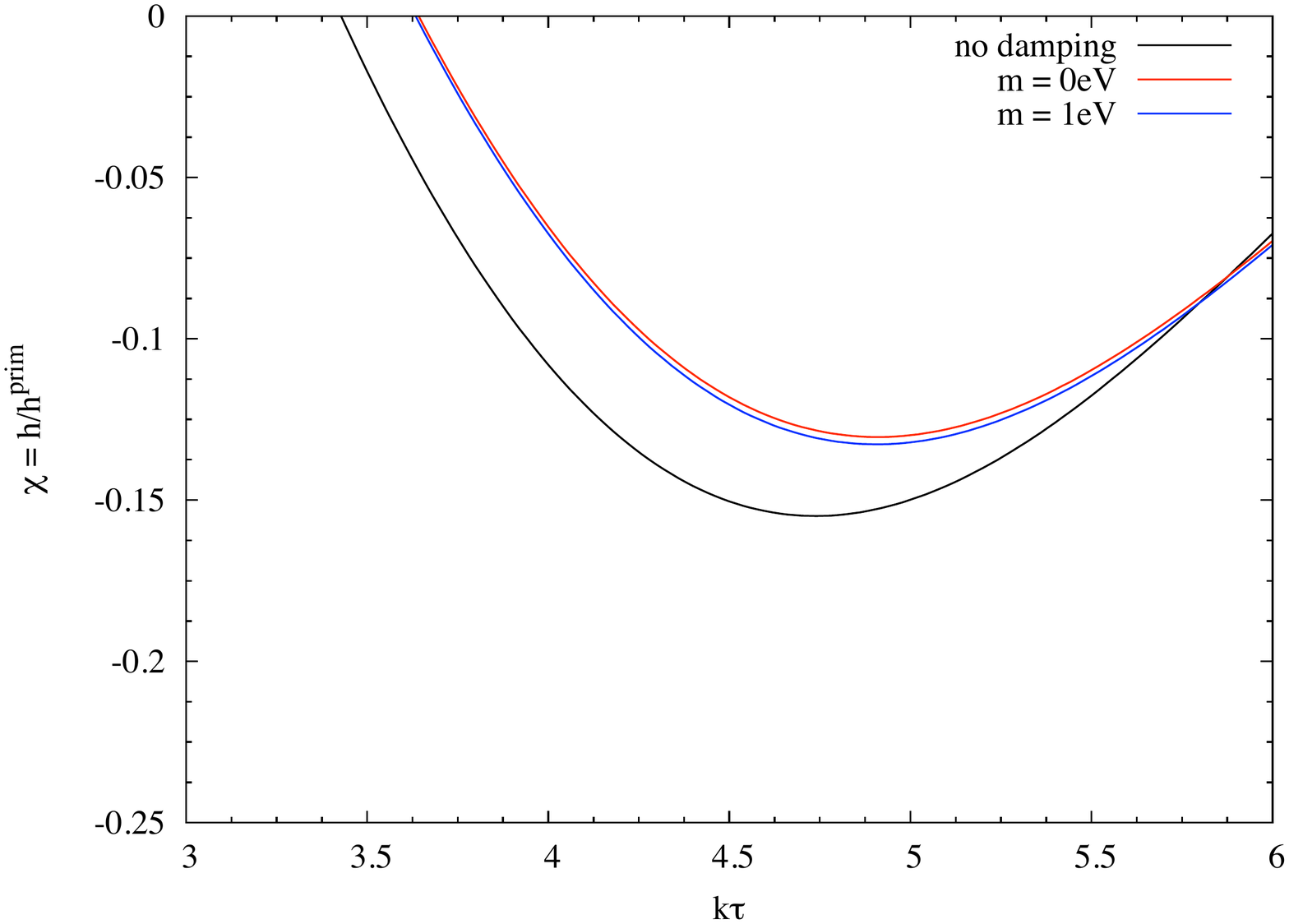}  
\includegraphics[width=3.5in]{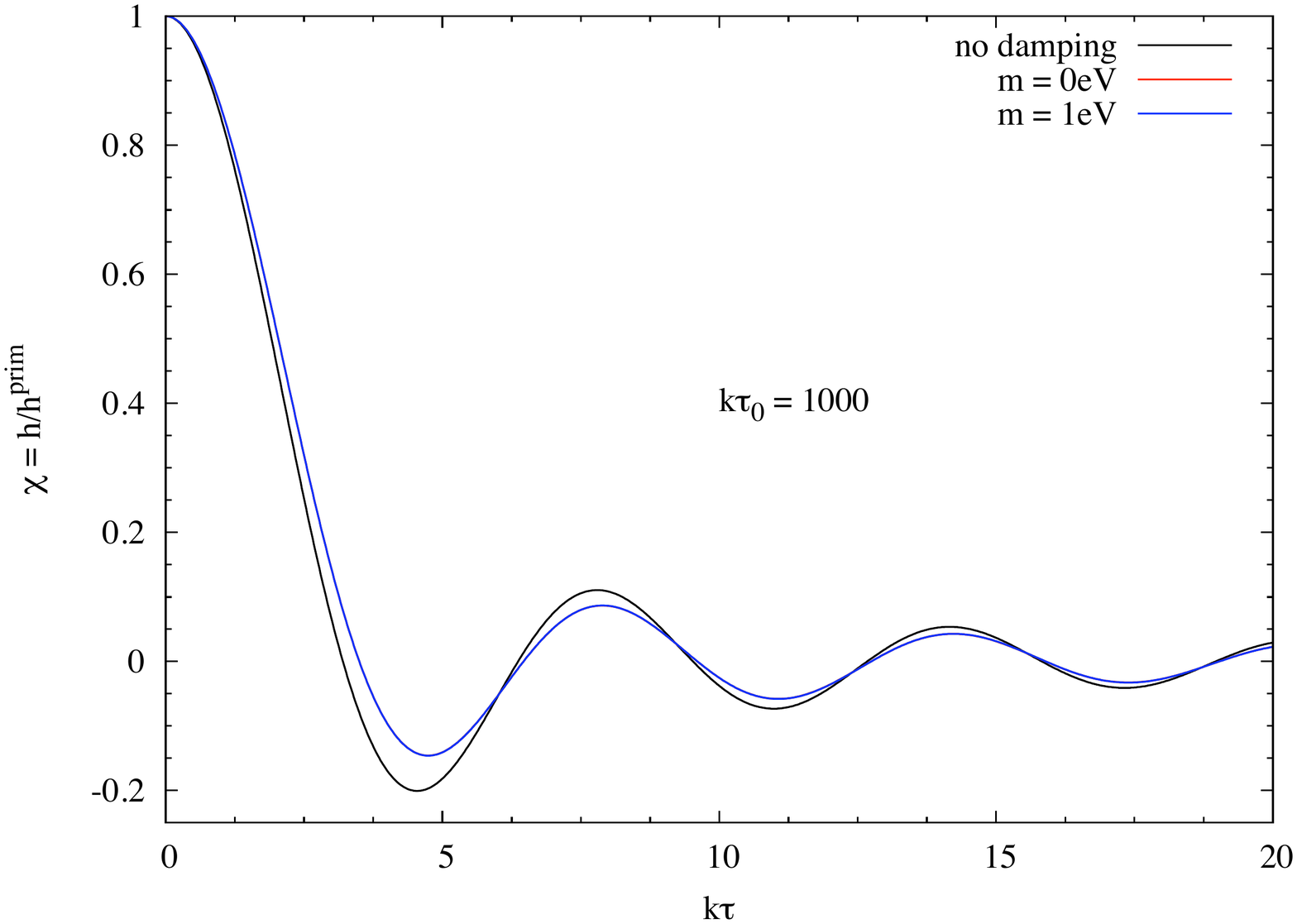} \includegraphics[width=3.5in]{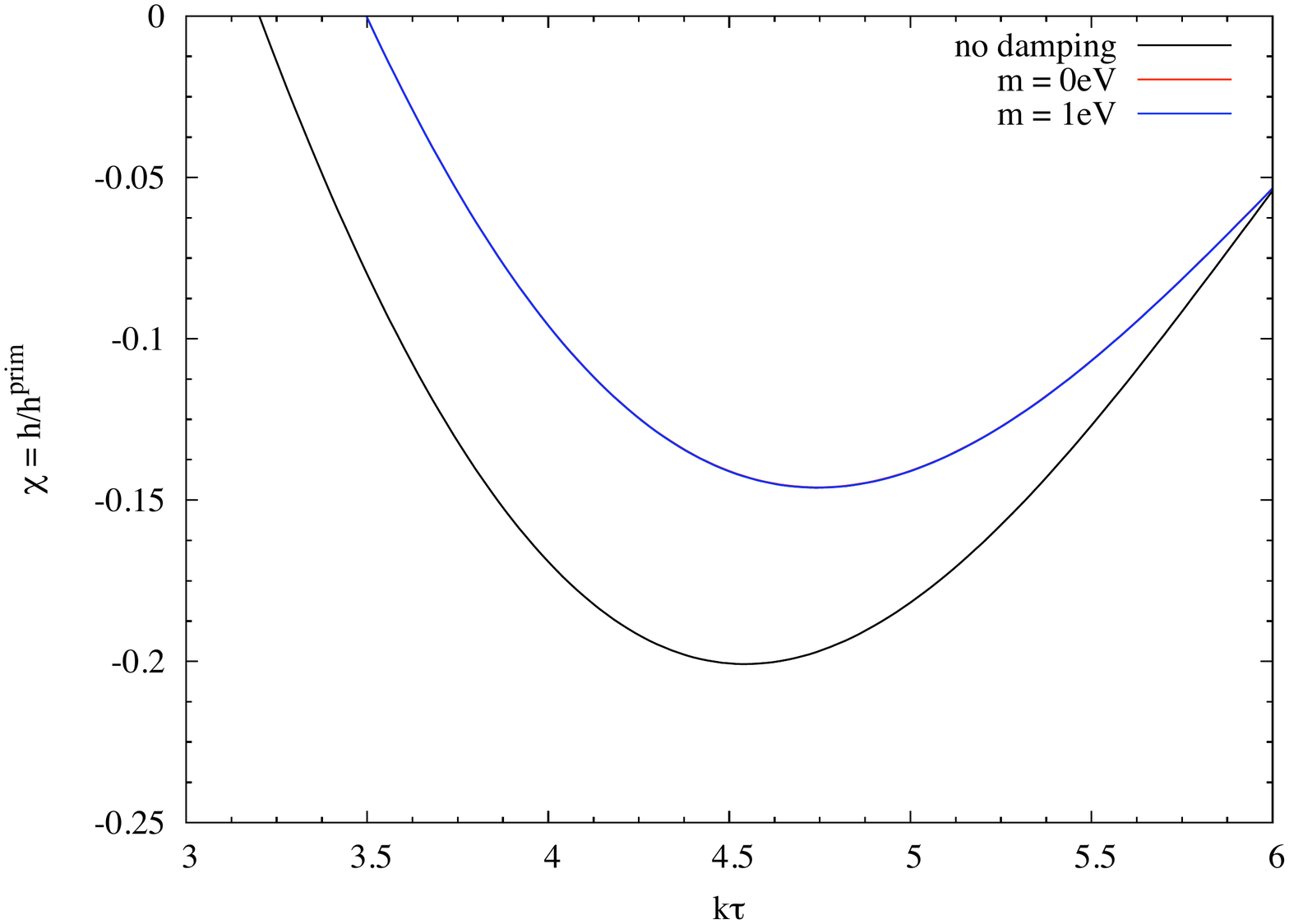}
 \caption{The $k$-dependence of the damping of an extra massive neutrino is demonstrated. The plots show that the damping is reduced for gravitational wave modes that enter the horizon as neutrinos are beginning to become non-relativistic. The damping is also less at later times when the neutrino energy fraction has been reduced due to redshifting. The region in each plot on the left around the first minima is zoomed in the adjacent plot on the right.}
   \label{fig:k_dependence}
   \end{figure}

In Fig.~\ref{fig:k_dependence}, we plot the damping from $3$ massless and a $1~{\rm eV}$
neutrino and compare it to the case of $4$ massless neutrinos. In doing so, we have adjusted $a_{\rm eq}$ to $N_{\rm eff}=4$ for both cases, which is a good approximation since neutrino masses of $O(1)~{\rm eV}$ are largely relativistic at matter-radiation equality.  Note that a cosmological scenario with a $1~{\rm eV}$ neutrino and $N_{\rm eff}=4$ is consistent with the current Planck data \cite{Planck:2013}. 

For $k\tau_{0}=100$, the ratio between the minima of $3$ massless plus $1~{\rm eV}$ and homogeneous ($undamped$) case is 0.94, and for the $4$ massless vs homogeneous case is 0.92, a difference of order $20 \%$. For $k\tau_{0}=200$, the difference in damping is of order $15 \%$.  And finally, for $k\tau_{0}=1000$ the difference in damping is now only $7 \%$ .

\begin{figure}[h!]
\includegraphics[width=0.8\textwidth]{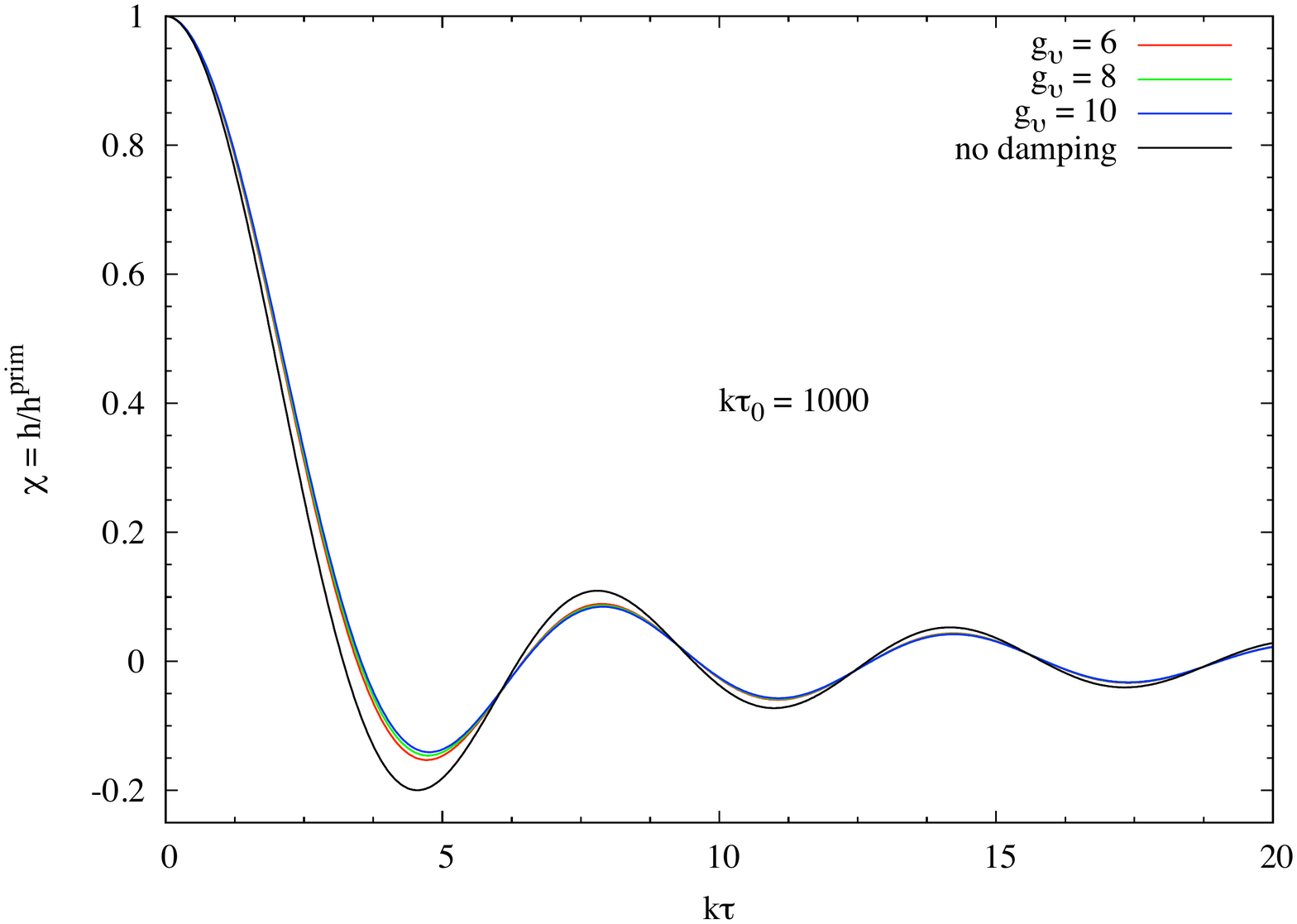}
\caption{The damping effect of extra massless neutrino species is shown. Each neutrino species has 2 degrees of freedom, thus $g_{\nu} = 6$, $8$, and $10$ correspond to 3, 4, and 5 neutrinos.  \label{fig:extraspecies}}
\end{figure} 

Similarly the effect of additional massless degrees of freedom is to increase the damping of gravitational waves.  
As seen in Fig. \ref{fig:extraspecies} this effect varies slightly with conformal time, $\tau$. 
We can compare the effect of extra species for example at the first minima. For the undamped case, this minima occurs at $u=4.54$ independent of $N_{\rm eff}$. However, including GW damping through free-streaming, this minima shifts to $u=4.72,\ 4.74,\ 4.76$ ($N_{\rm eff}=3,\ 4,\ 5$ respectively). With respect to homogenous case, the mode amplitude at the minima is $76.5\%$, $73.1\%$ and $70.5\%$ as large for $N_{\rm eff}=3,\ 4,\ 5$ as large, respectively. Thus, tensor modes are damped more, with increasing $N_{eff}$, as expected.

Since the identity of the source of any possible extra degrees of freedom is currently unknown, one may want to expand the realm of possibilities to include bosonic degrees of freedom.  As expected on the basis of number of degrees of freedom, and hence $N_{\rm eff}$, the damping due to a single boson is less by about $19\%$, than for that of a single, massless neutrino species.  Two bosonic degrees of freedom are virtually indistinguishable from a single neutrino.

\begin{figure}[h!]
\includegraphics[width=0.8\textwidth]{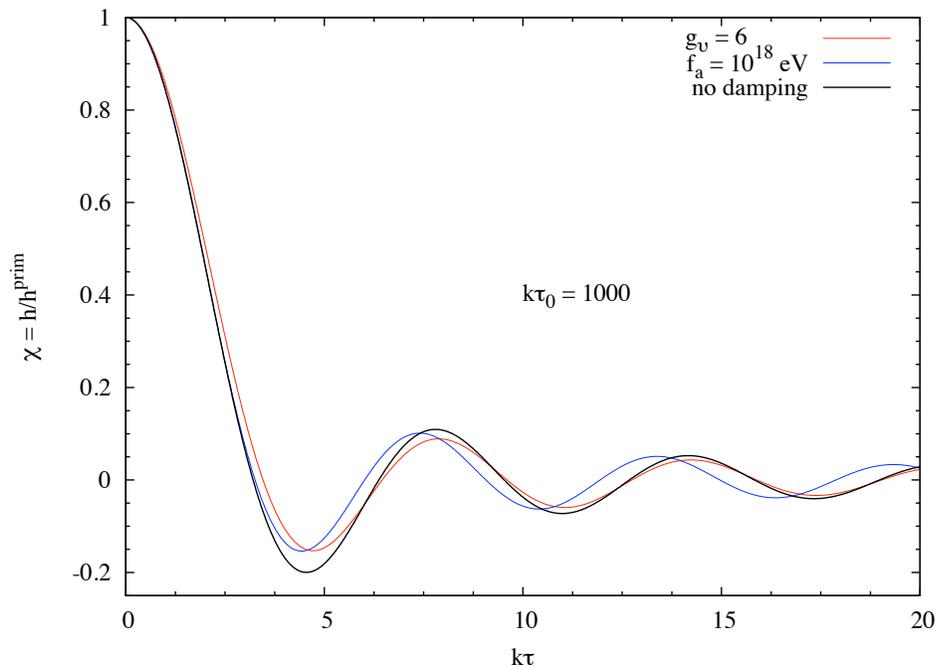}
\caption{The effect of axions produced by axionic strings.\label{fig:axions}}
\end{figure} 

In Fig.~\ref{fig:axions}, we examine the damping of gravitational waves caused by relativistic axions. We compare the results for axions with $f_{a}=10^{9}~{\rm GeV}$ versus $3$ massless neutrinos
since relativistic axions have $26\%$ and
$3$ massless neutrinos constitute $10\%$ of critical density at last scattering.
With respect to the no-damping case, the mode amplitudes at the minima are damped by $76.5\%$ for $3$ neutrinos versus $77\%$ for axions. However, note that the minima for neutrinos is at $u=4.72$ but for axions it is at $u=4.42$. And as can be seen this phase shift persists throughout the time evolution. Thus, although axions damp the amplitudes by the same amount as neutrinos for these parameters, their phase shift is an important distinguishing feature.

We can understand this effect on physical grounds. The axion phase space distribution, Eq.~(\ref{eqaxion}), has an explicit time dependence that is not present in the thermal neutrino distribution function.  As a result the integral over time of the anisotropic stress, which produces the damping, is modulated compared to the neutrino case, and hence modulates the resulting $k$-dependent damping of gravitational waves.

This phase difference will have an observational impact on the damping of CMB B-modes. 
Recall that it is $\dot{\chi}$ that enters into the Boltzmann equation for the temperature perturbations \cite{Weinberg:book}. Following \cite{Weinberg:2003ur}, we expect all tensor multipole coefficients to depend on $\chi(u)$ only through a factor of $|\chi'(u_{\rm LSS})|^2$, where $u_{\rm LSS}=(1+z_{\rm EQ})/(1+z_{\rm LSS})$ is the value of $u$ at the last scattering surface (LSS). 
We take $z_{\rm LSS}=1089$ and convert into $u_{\rm LSS}$ using Eq.~(\ref{scalefactor}) and $\Omega_{M}=0.3$. Moreover, we expect the dominant contribution to multipole $l$ in the CMB will come from wavenumber, $k\approx a_{\rm LSS}l/d_{\rm LSS}$ \cite{Weinberg:2003ur} where $a_{\rm LSS}$ is the scale factor at the surface of last scattering and $d_{\rm LSS}$ is the angular diameter distance of the surface of last scattering. Using numerical values of $a_{\rm LSS}$ and $d_{\rm LSS}$ we get
\begin{eqnarray}
l=0.878u_{\rm LSS}.
\end{eqnarray}

\begin{figure}[h!]
\includegraphics[width=0.8\textwidth]{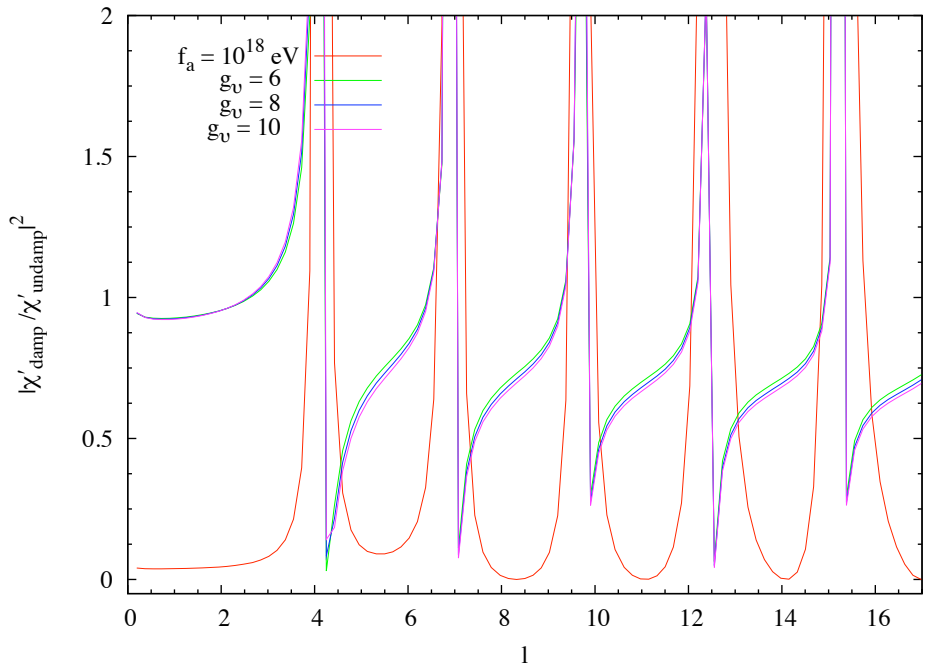}
\caption{The square of the ratio of the time derivative of damped modes to undamped modes which is useful for calculating the $B$ mode correlation function $C_{l}^{BB}$.}
\label{fig:power}
\end{figure} 

In Fig.~\ref{fig:power} we show the ratio of $\chi'^2$ for damped to undamped gravitational waves for axions and different numbers of massless neutrinos.  We have extracted this ratio at the surface of last scattering for several different low $l$ values. We expect the graph to look similar at higher
$l$ values but the computations at high $l$ become prohibitively expensive.
Both neutrinos and axions produce an oscillatory pattern in the damping 
but there is a phase shift between them. 
It is important to note that at certain $l$, ``damped'' gravitational waves can actually produce a larger signal than undamped waves by a factor of 
$2$ or more.  This surprising effect is due to that fact that for some $k\tau$ values there is 
actually a relative amplification caused by anisotropic stress, as can be seen from Figs. \ref{fig:k_dependence} and \ref{fig:extraspecies}, where the mode amplitude does not decrease 
as rapidly as in the undamped case.

\section{Conclusions}\label{conclusions}

The observation of a primordial gravitational wave spectrum would provide a direct window on physics of the very early universe.   As has been stressed, in order to extract as much cosmological information as possible from such a signal, one must be mindful of any phenomena which may alter the primordial signal.  One example of such a process is the damping of gravitational waves by free-streaming particles such as neutrinos. 

In this work we have generalized the formalism for deriving the effects of damping of gravitational waves due to anisotropic stress caused by free-streaming by deriving a general formula for the anisotropic stress as a function of mass and number of degrees of freedom, which should be useful for calculating the cosmological signature of possible additional non-standard model relativistic species.   

We find that for additional neutrino masses of current cosmological interest, the effects of non-zero mass on damping in comparison to the massless case is most pronounced for $k\tau_0 \approx 100-200$. For longer wavelength modes, that enter the horizon later, the damping is suppressed for all cases because the neutrino energy density is less significant.   In addition we have explored the possible impact of a relativistic axion background, as might be present due to radiation from axion strings.  While the overall damping produced by such a background could perhaps be comparable to that due to three standard model neutrinos, we find that their non-thermal phase space distribution will produce a possibly measurable phase shift in the damping signature. 

If a non-zero tensor B-mode contribution is observed in future CMB experiments, one might hope to use these results to help constrain new physics beyond the standard model.

\section{Acknowledgements}

We would like to thank Chiu-Man Ho for helpful discussions.  JBD recognizes support from the Louisiana Board of Regents and the NSF, and would like to thank the Research School of Astronomy \& Astrophysics at Australian National University for their hospitality while this paper was being completed. TV thanks the Institute for Advanced Study, Princeton, for hospitality. 
The work of LMK, SS, and TV is supported by the DOE at ASU.

\end{document}